\documentclass[preprint,11pt]{elsarticle}

\usepackage{amssymb}
\usepackage{soul}
\usepackage{xcolor}
\usepackage{booktabs}
\usepackage{multirow}
\usepackage{xcolor}
\usepackage{lipsum} 
\usepackage{afterpage}
\usepackage{float} 
\usepackage{amsmath}
\usepackage{caption}

\begin{document}

\begin{frontmatter}



\title{Adoption of AI-Assisted E-Scooters: The Role of Perceived Trust, Safety, and Demographic Drivers}



\author[inst1]{Amit Kumar}

\affiliation[inst1]{organization={William States Lee College of Engineering, University of North Carolina at Charlotte},
            addressline={9201 University City Blvd}, 
            city={Charlotte},
            postcode={28223}, 
            state={North Carolina},
            country={USA}}

\author[inst2]{Arman Hosseini}
\author[inst1]{Arghavan Azarbayjani}
\author[inst3]{Arsalan Heydarian}
\author[inst1]{Omidreza Shoghli \corref{cor1}}

\cortext[cor1]{Corresponding author. Email: \texttt{oshoghli@charlotte.edu}}

\affiliation[inst2]{organization={Systems and Information Engineering, University of Virginia},
            addressline={Olsson Hall}, 
            city={Charlottesville},
            postcode={22903}, 
            state={Virginia},
            country={USA}}

\affiliation[inst3]{organization={Civil and Environmental Engineering, University of Virginia},
            addressline={Olsson Hall}, 
            city={Charlottesville},
            postcode={22903}, 
            state={Virginia},
            country={USA}}
\begin{abstract}
E-scooters have become a more dominant mode of transport in recent years. However, the rise in their usage has been accompanied by an increase in injuries, affecting the trust and perceived safety of both users and non-users. Artificial intelligence (AI), as a cutting-edge and widely applied technology, has demonstrated potential to enhance transportation safety, particularly in driver assistance systems. The integration of AI into e-scooters presents a promising approach to addressing these safety concerns. This study aims to explore the factors influencing individuals’ willingness to use AI-assisted e-scooters. Data were collected using a structured questionnaire, capturing responses from 405 participants. The questionnaire gathered information on demographic characteristics, micromobility usage frequency, road users' perception of safety around e-scooters, perceptions of safety in AI-enabled technology, trust in AI-enabled e-scooters, and involvement in e-scooter crash incidents. To examine the impact of demographic factors on participants' preferences between AI-assisted and regular e-scooters, decision tree analysis is employed, indicating that ethnicity, income, and age significantly influence preferences. 
To analyze the impact of other factors on the willingness to use AI-enabled e-scooters, a full-scale Structural Equation Model (SEM) is applied, revealing that the perception of safety in AI enabled technology and the level of trust in AI-enabled e-scooters are the strongest predictors. 
\end{abstract}




\end{frontmatter}


\section{Introduction}
Electric scooters (e-scooters) have emerged as a transformative mode of mobility experiencing a remarkable rise globally, becoming a prominent feature in urban transportation networks \cite{shaheen2020sharing}. 
Their adoption has been particularly evident in US, where micromobility trips increased from 321,000 in 2010 to 157 million in 2023, with shared e-scooters playing a key role in this growth \cite{nacto2024shared}.
E-scooters offer a convenient and sustainable alternative to traditional urban transit, providing essential first- and last-mile connectivity \cite{burt2023scooter}.
Their ease of use has made them a viable option for short-distance travel, often surpassing the utility of shared bicycle services \cite{hardt2019usage}.

However, several challenges hinder their integration into urban transportation systems, creating difficulties for municipal and regulatory agencies in addressing their impacts on urban spaces and road safety \cite{comi2022innovative}. 
One of the most pressing issues associated with the increasing use of e-scooters is safety. Infrastructure inadequacies exacerbate these challenges, as many cities lack dedicated lanes for e-scooters, forcing riders to share roads with vehicles and sidewalks with pedestrians \cite{chen2024impact}. The predominance of car-centric urban design often creates hazardous situations, with vehicles posing substantial risks to e-scooter users. A primary concern among riders is the threat of being struck by a moving vehicle or colliding with one \cite{sievert2023survey}.
Additionally, the absence of well-defined laws for e-scooter usage creates ambiguity, with regulations on parking zones, speed limits, and permitted riding areas varying across locations \cite{arun2021systematic}.
Riders are particularly vulnerable to traumatic injuries due to the lack of protective gear and the ergonomic design of e-scooters \cite{trivedi2019craniofacial}. 
The spontaneous nature of e-scooter usage discourages helmet use, as riders typically do not carry helmets for impromptu trips \cite{serra2021head}. 

The safety challenges of e-scooters are not limited to the e-scooter riders. Riding e-scooters on sidewalks has led to discomfort and crashes between riders and other vulnerable road users, particularly pedestrians. Accidents involving collisions with vehicles or infrastructure have raised liability issues and highlighted the inexperience of some riders \cite{kazemzadeh2023electric}. 
The speed differential between pedestrians and e-scooter users can result in heightened collision risks, as approximately 40\% of e-scooter trips occur within four feet of a pedestrian \cite{haworth2021comparing,feng2022estimating}. 

Although these challenges persist, significant advancements in AI and autonomous vehicle (AV) technologies provide a promising framework for addressing such issues. In particular, 
integrating cutting-edge sensors and AI-assisted driving technologies has shown potential to enhance safety. Advanced driver assistance systems, operating at various levels of autonomy, can identify and mitigate potential hazards, alert drivers to dangers, and assist in accident prevention \cite{greenblatt2015automated}. E-scooters can benefit from these technologies by incorporating AI-enabled systems such as real-time object detection \cite{chen2024performance}, collision avoidance \cite{li2023modeling}, and adaptive speed controls. 

Despite the numerous studies on e-scooter adoption \citep{mckenzie2020urban, rejali2021assessing}, research on trust and perceived safety has primarily centered on regular e-scooters. This has left the impact of AI integration into e-scooters on road users' perceptions under-explored \citep{samadzad2023factors, ari2024investigating}. There is limited understanding of the adoption of AI-assisted e-scooters and the factors that influence the willingness to use them.
This paper addresses these gaps by answering the following research questions: (1) What is the impact of sociodemographic characteristics on adoption of AI-enabled e-scooters? (2) How do individual perceptions, behaviors, and trust dynamics interact and shape the decision to adopt AI-enabled e-scooters?

\section{Literature Review} \label{sec:literature}

\subsection{Perceptions of Safety in E-Scooter Use: Rider and Non-Rider Perspectives}
Safety is a critical concern regarding e-scooter use, and many studies focused on injury and crash data to address these concerns. However, another equally important aspect is the perception of safety, including how both users and non-users view e-scooters in terms of safety. This includes factors such as riding behavior, interactions with other road users (such as pedestrians, cyclists, and drivers), and the general feeling of safety while riding.

From the riders' perspective, one of the most significant safety concerns is road infrastructure. Many riders report a low sense of safety when navigating major streets, largely due to the risk of collisions with vehicles. A particularly prominent concern is the possibility of encountering potholes, which can destabilize the rider and lead to crashes \cite{sievert2023survey}. Both riders and non-riders perceive riding e-scooters in vehicle lanes as unsafe \cite{pourfalatoun2023shared}, reflecting a shared apprehension about the risks posed by mixed-traffic environments.
The study by Tian et al. \cite{tian2022characteristics} revealed that, although sidewalks accounted for the majority of e-scooter-related injuries (44\%, primarily minor injuries), they are still perceived by riders as the safest type of road infrastructure. This perception likely stems from the reduced interaction with vehicles, which riders associate with a lower risk of severe collisions, despite the higher incidence of minor injuries on sidewalks.

Several studies highlighted the safety concerns that e-scooters pose from the perspective of other road users. Všucha et al. \cite{vsucha2023scooter} conducted surveys across five countries, finding that non-riders generally perceived e-scooter use as dangerous, with women and older individuals more likely to consider them unsafe.
Derrick et al. \cite{derrick2020perceptions} surveyed participants in Singapore to assess pedestrian safety concerns and support for an e-scooter ban. The findings showed that 64\% of respondents supported a ban on e-scooters, citing their danger to pedestrians. Key factors influencing this support included age, negative personal experiences, and social norms (e.g., family or peer opinions), and speeding was identified as the most common safety concern.
James et al. \cite{james2023pedestrians} reported that pedestrians and drivers felt less safe walking or driving around dockless e-scooters compared to bicycles. In another study, most pedestrians and cyclists perceived abandoned e-scooters on sidewalks as hazards to other road users \cite{burt2023scooter}.

Although previous studies examined safety perceptions of regular e-scooters, the perception of safety in AI-integrated e-scooters remains unexplored. Further studies are required to assess how safety perceptions of AI enabled technology and road users' views on e-scooters influence overall safety perception and trust in AI-enabled e-scooter. 

\subsection{Perception of Safety and Trust in AI-assisted Vehicles}
The level of trust in AI-assisted vehicles has been a significant topic of interest since the advent of autonomous vehicles (AVs). Public concerns about AVs are often amplified by reports of accidents involving these vehicles, which can further hinder their acceptance. Previous experiences with AVs play a crucial role in shaping public attitudes toward this technology. Individuals with prior exposure to AVs tended to exhibit more positive perceptions and a greater willingness to adopt them compared to those without such experiences \cite{othman2021public}. Previous studies examined the impact of sociodemographics on trust in AI-assisted vehicles. One study revealed that individuals aged 36 to 65 expressed greater apprehension and resistance toward driving autonomous vehicles compared to both younger individuals aged 18 to 35 and older individuals aged 65 and above \cite{thomas2020perception}. In other study, older people generally held more pessimistic views about AVs \cite{othman2021public}. Gender differences also emerged, with males expressing more positive attitudes toward AVs than females.
In terms of education, individuals with a university degree (Bachelor's, Master's, or PhD) exhibited lower levels of concern regarding accident liability and system failures in autonomous vehicles compared to those without a degree \cite{thomas2020perception}. This finding aligns with another study, which found that people with higher education levels are more likely to have a positive view of AVs than those with lower levels of education \cite{othman2021public}. Interestingly, individuals in countries with lower GDP levels tended to have more positive attitudes toward AVs compared to those in medium- or high-GDP countries\cite{othman2021public}.



Pyrialakou et al. \cite{pyrialakou2020perceptions} examined safety perceptions among vulnerable road users interacting with AVs. Their findings highlighted gender differences, with females feeling less safe around AVs than males. Among various activities, cycling near AVs was perceived as the least safe, followed by walking and then driving. The study also found that feeling safe near AVs was positively associated with reduced concerns about threats such as hacking or terrorism. Additionally, direct experience with AVs significantly improved pedestrian safety perceptions.

Previous studies primarily examined the trust and safety perceptions in autonomous vehicles, with limited attention to other AI-assisted modes of transport. To the best of our knowledge, no study has specifically investigated the perception of trust and safety in AI-assisted e-scooters, either from the perspective of riders or other road users. This study explores the level of trust in AI-assisted e-scooters and its influence on the willingness to adopt this emerging system.

\subsection{Adoption of E-scooter}
Previous studies explored factors influencing the adoption of electric scooters. 
Sanders et al. \cite{sanders2020scoot} investigated the willingness to use e-scooters in Arizona, U.S., revealing significant gender-based differences in barriers to usage, especially regarding safety concerns. Moreover, African American and non-white Hispanic respondents were more likely than non-Hispanic white respondents to express an intention to try e-scooters. Nikiforiadis et al. \cite{nikiforiadis2021analysis} examined shared e-scooter usage in Thessaloniki, Greece, and found that females were less inclined to use e-scooters compared to males.
Teixeira et al. \cite{teixeira2023barriers} conducted a survey across five European capital cities to explore the barriers preventing non-users from adopting e-scooters. The results implied that these obstacles are largely external and infrastructural, including the convenience of alternative transport modes, safety concerns about riding in traffic and inadequate road conditions. Similarly, She et al.\cite{she2017barriers} identified barriers to the widespread adoption of electric scooters and found that younger individuals showed more positive attitudes toward using them compared to older generations.

Structural Equation Modeling (SEM), often based on the extended Technology Acceptance Model (TAM), is a widely used approach for analyzing factors influencing e-scooter adoption\cite{ari2024investigating,samadzad2023factors,javadinasr2022eliciting}. Ari et al. \cite{ari2024investigating} found that social influences and enjoyment were the most significant predictors of perceived ease of use and perceived usefulness. Samadzad et al. \cite{samadzad2023factors} highlighted that perceived usefulness, trust, and subjective norms are key factors shaping the adoption and willingness to use shared e-scooters. Javadinasr et al. \cite{javadinasr2022eliciting} focused on factors driving the continuous use of e-scooters in Chicago, showing that perceived usefulness is the most influential factor, followed by perceived reliability.

Despite significant research on the adoption of regular e-scooters, no studies, to the best of our knowledge, have specifically examined the willingness to use AI-assisted e-scooters. Key questions remain unresolved, such as identifying the profiles of individuals who prefer AI-assisted e-scooters over regular ones and uncovering the factors that influence the willingness to adopt this new system.

\section{Model Structure and Hypothesis}
To address the research questions, we developed two models: (1) a Decision Tree Classifier to find patterns in sociodemographic features that shape participants' preference between regular vs AI-assisted e-scooters. (2) Structural Equation Modeling (SEM) to detect factors significantly impacting the willingness to use AI-assisted e-scooters. 
 
\begin{figure}[!ht]
    \centering
    \includegraphics[width=0.8\linewidth]{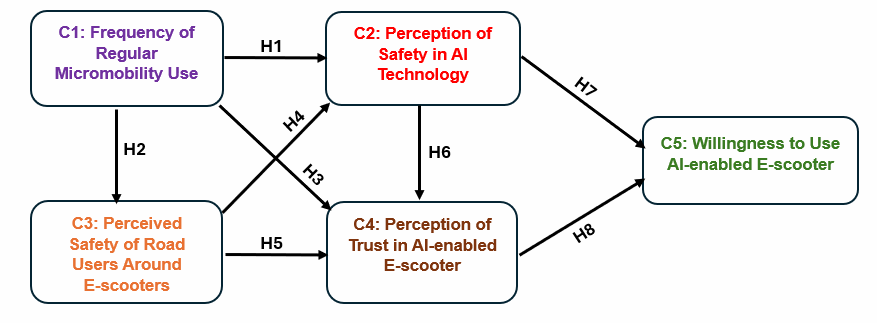}
    \caption{The proposed model to assess willingness to use AI enabled driving assistance technology}
    \label{SEM modelZ}
\end{figure}

The SEM framework was designed to analyze the causal relationships between latent variables, enabling the assessment of both direct and indirect effects within a theoretical model. By incorporating structural paths, we aimed to evaluate the underlying mechanisms shaping participants’ attitudes and behaviors toward AI-assisted e-scooters. Specifically, the model considers five key constructs: (1) Frequency of regular micromobility use, (2) Perceived safety of road users around e-scooters, (3) Perception of safety in AI-technology, (4) Perception of trust in AI-enabled e-scooter, (5) Willingness to use AI-enabled e-scooter.
Figure \ref{SEM modelZ} illustrates the proposed constructs of our model along with their hypothesized relationships. The rationale for selecting these factors, as well as the detailed explanation of the proposed hypotheses, are provided in the following subsections.

\subsection{Frequency of Regular Micromobility Use}
The frequency of micromobility use varies significantly across different urban environments and demographic groups. In dense city centers, many people use bicycles, e-scooters, and e-bikes multiple times per week or even daily for commuting and short trips, particularly in areas with well-developed cycling infrastructure and favorable weather conditions \citep{blazanin2022scooter,bachand2012better, xu2019analysis}.
At the same time, the rapid integration of artificial intelligence (AI) technologies into micromobility solutions, including AI-assisted e-scooters, holds the potential to significantly influence urban mobility landscapes in the near future \cite{alp2025micromobility}. AI-powered features such as real-time obstacle detection, adaptive speed control, and collision detection aim to enhance both the safety and trust of vulnerable road users (VRUs). However, perceptions of these advancements which are key in technology acceptance are underexplored.

Users with frequent engagement in regular e-scooters and shared micromobility may develop a more nuanced understanding of road risks, operational dynamics, and vehicle behavior, potentially influencing their perception of AI-enhanced safety features. Similarly, regular micromobility users may develop adaptive strategies for interacting with e-scooters, fostering greater acceptance. However, non-users or infrequent users may be more likely to associate e-scooters with concerns such as pedestrian safety conflicts, regulatory ambiguity, and infrastructure inadequacy. This suggests that prior micromobility engagement may influence attitudes toward AI-enabled e-scooters \cite{jevinger2024artificial}. 

Trust in AI-powered mobility solutions is a critical determinant of adoption intention \cite{kuberkar2020factors}. Frequent users may trust AI for better navigation and safety, while less experienced users may doubt its ability to handle urban complexities and non-motorized interactions.
Based on this information, we propose the following hypothesis:

\begin{itemize}
    \item H1: Frequency of regular micromobility use positively influences road users' perception of safety in AI enabled technology.
    \item H2: Frequency of regular micromobility use positively influences perceived safety of road users around e-scooters.
    \item H3: Frequency of regular micromobility use positively influences perception of trust in AI-enabled e-scooter.
\end{itemize}

\subsection{Perceived safety of road users around e-scooters}
Public attitudes toward regular e-scooters may shape how users evaluate AI-driven E-Scooters. 
Perceived safety of road users around e-scooters aims to evaluate how different road users perceive their safety when e-scooters are in their proximity. As a result, we need to assess the safety perception of other road users, incluing: pedestrians, cyclists, e-bike riders, vehicle drivers, and e-scooter riders, in respone to presence of regular e-scooters.
Positive experiences with traditional e-scooters could lead to greater openness toward AI enhancements, while negative perceptions may result in hesitation or resistance to adopting AI-powered alternatives, and lack of trust in AI-enabled technologies.
Based on this construct model, we propose the following hypothesis: 
\begin{itemize}
    \item H4: Perceived safety of road users around e-scooters positively influences perception of safety in AI-enabled technology.
    \item H5: Perceived safety of road users around e-scooters positively influences perception of trust in AI-enabled e-scooter.
\end{itemize}

\subsection{Perception of Safety in AI-enabled technologies}
Perception of Safety in AI-enabled technologies assesses individuals' confidence in the ability of AI systems in vehicles to operate securely and minimize risks in real-world scenarios. It explores users' beliefs about the safety measures, reliability, and decision-making accuracy of AI-powered solutions across different applications, particularly in mobility and transportation. Previously, Jevinger et al.\cite{jevinger2024artificial} used machine learning technique to present a case for AI improving public transport safety while Azarbayjani \cite{azarbayjani2024trust} investigated the trust and perceived safety of vulnerable adult road users towards regular and AI-enabled E-scooters. A strong perception of safety is essential for building user trust, as it alleviates concerns about potential errors, malfunctions, or unforeseen hazards. This construct is crucial for evaluating public acceptance and safety in AI enabled technologies, particularly because concerns about system failures, unpredictability, or decision- making in AI algorithms may erode trust, leading to reluctance in adoption.
Based on this construct model, we propose the following hypothesis: 
\begin{itemize}
    \item H6: Perception of safety in AI-enabled technologies positively influences perception of trust in AI-enabled e-scooter.
    \item H7: Perception of safety in AI-enabled technologies positively influences willingness to use AI-enabled e-scooter.
\end{itemize}

\subsection{Perception of Trust in AI-Enabled E-Scooter}
Trust is the most relevant psychological state which is derived from the willingness to accept vulnerability and has been identified as a key factor in the adoption and use of technology \citep{rousseau1998not,hohenberger2017not}. 
The perception of trust in AI-enabled e-scooters construct focuses on evaluating users' perception of the ability of the AI technology integrated into e-scooters to handle unpredictable situations, functionality, and maneuvering. It examines how well users trust the AI to enhance safety, provide accurate navigation, and ensure smooth operation in various environments. Xu et al. \cite{xu2019analysis} identified that trust has a strong and positive correlation with both perceived usefulness and ease of use. Trust is crucial for fostering widespread adoption of AI-enabled e-scooters, as it may influence users' willingness to use AI-enabled e-scooters. 
Based on this construct model, we propose the following hypothesis: 
\begin{itemize}
    \item H8: Perception of trust in AI-enabled e-scooter positively influences willingness to use AI-enabled e-scooters.
\end{itemize}

\section{Method} \label{sec: Method}
The research methodology of this study involved designing and distributing an online survey to collect data across the United States. The collected data were analyzed using decision tree classification to explore the impact of sociodemographic factors on preferences for AI-assisted e-scooters, and structural equation modeling (SEM) to examine complex relationships between perceived safety, trust, and willingness to use AI-enabled e-scooters. In the following sections, we explain the detailed methodology used for survey design, distribution, participant selection, data analysis, and model development.

\subsection{Survey Design}\label{sec:design}
To collect data for the model outlined in the previous section and test the hypotheses, we developed a survey questionnaire aligned with the model's structure and hypotheses. The full list of questions used for hypothesis testing can be found in Table \ref{tab:reliability-validity}. The survey is organized into four sections as follows.
\begin{enumerate}
\item Perceived safety of road users around e-scooters:
The survey began by exploring participants' interactions with e-scooters, tailored to their role as road users—pedestrians, car drivers, bike riders, e-bike riders, or e-scooter riders. Respondents were asked about their perceived safety of e-scooters, while those with prior e-scooter experience provided more detailed information, including their frequency of use and any previous accident experiences. These questions are listed from question 1 to question 8 under the Survey Questions in Table \ref{tab:reliability-validity}.

\item Perceptions of safety in AI-enabled technologies:
To assess participants' perceptions of safety in AI-enabled technology, the survey included questions examining their overall safety perceptions, comparative views on autonomous versus human-driven vehicles, and their evaluation of AI-assisted e-scooter features. Responses were collected using both open-ended and Likert scale formats to capture nuanced perspectives on safety in AI-enabled technology. The relevant questions in this section are listed from question 9 to 11 in Table \ref{tab:reliability-validity}.

\item Perceived trust in AI-enabled e-scooters:
In this section, participants, regardless of their prior experience with e-scooters, were presented with a conceptual scenario involving e-scooters equipped with AI systems. They were asked to share their perceptions of trust and their preferences for how such AI systems could assist them with notifications. The questions were designed to assess participant's confidence in the system's ability to enhance ride safety, handle unexpected situations, and operate reliably without malfunctioning. The appropriate questions for this section are listed from question 12 to 14 in Table \ref{tab:reliability-validity}.

\item Demographic information:
The final section included demographic questions to collect information on participants' personal characteristics, such as age, gender, race, ethnicity, education level, income, and geographic location. These details were essential for understanding the composition of the respondent pool and categorizing populations within the study. By analyzing demographic factors, the study aimed to develop a comprehensive analysis of trends and patterns across different population segments.
\end{enumerate}

\subsection{Survey Distribution} \label{sec:distribution}
The online survey was administered using Qualtrics, a cloud-based platform for survey design and distribution. 
Participants accessed the survey using an active link generated by the platform, which was compatible with a wide range of devices, ensuring accessibility and convenience. The survey was distributed through multiple channels, including email, pamphlets in transit systems including buses and light rails,  social media platforms, public space gatherings including weekend farmer's markets, public parks, and Thanksgiving parade, university classes at the University of North Carolina at Charlotte and the University of Virginia, and campus communications. Each distribution was accompanied by a brief description of the study’s purpose, assurance of confidentiality, and clear instructions for participants. To encourage participation, respondents were offered the option to receive a \$3 online gift card incentive upon completion of the survey. Ethical approval for the study was granted by the Institutional Review Boards of the University of North Carolina Charlotte (IRB-24-0118) and the University of Virginia (IRB-6120).
Data collection spanned a 12-month period, from November 2023 to November 2024. The survey, with an estimaed completion time of 15 minutes, targeted a diverse population of road users across the United States, including both e-scooter riders and non-riders.
 
\subsection{Participants}
Participation was limited to individuals aged 18 or older who resided in the United States. A total of 405 valid responses were collected, comprising 188 e-scooter users (46\%) and 218 non-users (54\%). 
The majority of respondents (56\%) were young adults aged between 18 and 25, which aligns with the target population of e-scooter users.
A detailed distribution of age ranges is provided in Table \ref{tab:demographics}.

Regarding gender identity, 233 (58\%) participants identified as male, 162(40\%) as female, 2 as non-binary, and 8 (2\%) preferred not to disclose their gender. The racial distribution of the sample was as follows: (42\%) identified as White, (24\%) as Asian, (11\%) as Hispanic or Latino, (9\%) as African American,(9\%) as Middle Eastern, (2\%) as others, and (2\%) of participants preferred not to disclose their ethnic background. A summary of the respondents' demographics is presented in Table \ref{tab:demographics}.

\begin{table}[H]
\centering
\scriptsize 
\setlength{\tabcolsep}{3pt} 
\renewcommand{\arraystretch}{1} 
\caption{Demographic Distribution of Survey Respondents}
\label{tab:demographics}
\resizebox{0.7\textwidth}{!}{ 
\begin{tabular}{llrr}
\toprule
Variable & Category & Count & Percentage (\%) \\
\midrule
Gender Identity & Female & 162 & 40\% \\
 & Male & 233 & 58\% \\
 & Non-binary & 2 & 1\% \\
 & Prefer not to disclose & 8 & 1\% \\
\midrule
Age & 18--21 & 113 & 28\% \\
 & 22--25 & 112 & 28\% \\
 & 26--30 & 51 & 13\% \\
 & 31--35 & 39 & 10\% \\
 & 36--40 & 31 & 8\% \\
 & 41--50 & 35 & 9\% \\
 & 51 or older & 24 & 3\% \\
\midrule
Racial Background & White/Caucasian & 170 & 42\% \\
 & Asian & 98 & 24\% \\
 & Hispanic/Latino & 44 & 11\% \\
 & Middle Eastern & 35 & 9\% \\
 & African American & 35 & 9\% \\
 & Native American & 2 & 1\% \\
 & Other & 10 & 2\% \\
 & Prefer not to disclose & 11 & 2\% \\
\midrule
Education Level & High school graduate & 99 & 25\% \\
 & Associate degree & 49 & 12\% \\
 & Bachelor's degree & 115 & 28\% \\
 & Master's degree & 94 & 23\% \\
 & Advanced degree & 47 & 12\% \\
\midrule
Income & Less than \$20,000 & 67 & 17\% \\
 & \$20,000 - \$50,000 & 87 & 21\% \\
 & \$50,000 - \$100,000 & 76 & 19\% \\
 & \$100,000 - \$150,000 & 68 & 16\% \\
 & \$150,000 - \$200,000 & 32 & 8\% \\
 & More than \$200,000 & 30 & 7\% \\
 & Prefer not to disclose & 45 & 11\% \\ 
\bottomrule
\end{tabular}
}
\end{table}

Regarding the highest level of education, 28\% of participants held a bachelor's degree, while 25\% reported having a high school diploma or equivalent. Additionally, 23\% had obtained a master's degree, 12\% held an associate degree, 11\% possessed a doctorate, and 1\% reported holding a professional healthcare degree. 

Concerning household income, 17\% of participants reported annual incomes below USD 20,000, 21\% fell within the USD 20,000– 50,000 range, and 19\% reported incomes between USD 50,000 –100,000. Additionally, 16\% were in the USD 100,000–150,000 bracket, 8\% in the USD 150,000 –200,000 range and 7\% indicated annual incomes exceeding USD 200,000. The remaining 11\% preferred not to disclose their income. Further details on education and income are provided in Table 1. 

\subsection{Decision Tree Classification Approach for Evaluating the Impact of Sociodemographic Features}
Decision trees are among the most widely used classification methods due to their simplicity, interpretability, and effectiveness in handling large datasets. 
As a nonparametric and nonlinear approach, decision trees do not rely on assumptions about data distribution or linearity, making them highly adaptable to various data types while maintaining satisfactory accuracy \cite{priyam2013comparative, de2013decision}.
This section explains the process of designing the decision tree classifier to analyze the influence of sociodemographic characteristics on their preferences for choosing between regular and AI-assisted e-scooters both of which were priced identically. 
Participants were presented with three options to express their choice: (a) I will definitely choose the AI-assisted e-scooter, (b) I am uncertain about the decision, and (c) I would prefer the regular e-scooter. The responses were analyzed in relation to various demographic factors, including age, gender, race, education level, and income, to identify any potential correlations. A classification tree analysis drawing from the methodology developed by Nikiforiadis et al.  \cite{nikiforiadis2021analysis} was conducted utilizing Python and scikit-learn library to examine these relationships. The distribution of choices among participants revealed that 205 observations were recorded for AI-assisted e-scooters, while 104 participants indicated they were unsure, and 93 observations were noted for regular e-scooters.


The class imbalance posed a potential bias for the decision tree model, as it could lead to a dominant class skewing the results. While methods such as undersampling and oversampling could address this issue, they come with their own drawbacks, such as the loss of valuable data or the creation of synthetic data. To mitigate this, the "Not sure" and "Regular e-scooter" categories were combined into a single class labeled "Not AI-assisted." This class represents participants who either chose the regular e-scooter or were uncertain about their decision, effectively consolidating non-AI-assisted choices into one category. 

The data was split into a training set (80\%, 321 data points) and a test set (20\%, 81 data points). To fine-tune the hyperparameters, a grid search cross-validation \cite{bramer2007avoiding} was performed, testing nine different combinations of parameters, including maximum depth, minimum samples split, and minimum samples leaf. Two criteria, Gini and entropy, were applied to evaluate the splits, with accuracy used as the scoring metric. The best-performing hyperparameters, yielding an accuracy of approximately 65.3\%, were selected for training the decision tree model. The result of decision tree analysis is provided in section \ref{sec: Result}.
 
\subsection{Structural Equation Modeling (SEM)}
SEM is a comprehensive statistical approach that combines factor analysis and multiple regression to analyze complex relationships among observed and latent variables \cite{ullman2012structural}.
The technique involves building a hypothesized model, estimating parameters, and evaluating model fit using indices such as the Root Mean Square of Approximation, Comparative Fit Index, and Tucker Lewis Index. A well-specified SEM provides valuable insights into the structural relationships and underlying mechanisms within data \cite{kline2023principles}. 
The SEM based approach used in the research aims to fit a theoretical model to observed data, making it particularly suitable for confirmatory research. SEM assumes that constructs are common factors and estimates the model accordingly, using a statistical model to analyze correlations between dependent and independent variables, as well as the hidden structures \cite{SmartPLS4}. The method employs sophisticated algorithms, such as maximum likelihood estimation, to simultaneously calculate model coefficients using all available information from the observed variables. This comprehensive approach allows researchers to test hypotheses and assess the consistency of their theoretical models with empirical data, providing valuable insights into causal relationships and latent constructs.
In this study, the full SEM was developed using IBM SPSS AMOS, it considered the simultaneous assessment of the impact of all the constructs, C1: 'Frequency of regular micromobility use', C2: 'Perceived safety of road users around e-scooters', C3: 'Perception of safety in AI-enabled technologies.', and C4: 'Perception of trust in AI enabled e-scooter', on C5: 'Willingness to Use AI enabled e-scooter.'

\subsubsection{Data Screening and Preparation}
Common Method Bias (CMB) may arise due to artificial correlations between variables, potentially distorting parameter estimates. This bias may stem from controllable and uncontrollable factors, including respondent behavior, questionnaire design, and survey administration conditions. To mitigate CMB in this study, survey questions were structured for clarity and comprehension. Additionally, items measuring the same construct were mixed together so that they were not consecutive in order to avoid false and strong correlations between variables.


Furthermore, the data screening in Structural Equation Modeling (SEM) involved identifying and addressing missing data issues using mean data imputation, outlier detection, and non-normality to ensure the accuracy and reliability of results. Assessing data quality for indicators within the same construct requires evaluating both, reliability, and validity. Key measures of reliability and convergent validity include Cronbach’s Alpha (CA), Composite Reliability (CR), and Average Variance Extracted (AVE). A Cronbach’s Alpha between 0.6 and 0.7 is considered acceptable \cite{eisinga2013reliability}, while an AVE of 0.5 or higher and CR of at least 0.6 indicate satisfactory construct validity \citep{fornell1981evaluating, hair2019use, diba2023autonomous}.The CA and CR was found to be more than 0.6 for all the latent variables with the exception of C1 for which the values were 0.45 and 0.46 respectively.

To ensure the quality of data across different constructs, discriminant validity was assessed using the Heterotrait-Monotrait (HTMT) ratio and the Variance Inflation Factor (VIF). Additionally, row-wise data quality was examined using SPSS and Microsoft Excel, focusing on response patterns to identify unengaged respondents. Participants who provided identical ratings (e.g., consistently selecting "5" across all items) were flagged as unengaged. In self-reported Likert scale surveys, particularly lengthy ones, respondents may exhibit response bias by selecting the same rating for all questions, regardless of item phrasing. To mitigate this issue, responses with a mean value below 0.5 or above 4.5 on the 5-point Likert scale were excluded. Furthermore, cases where the standard deviation was less than 0.25—indicating a lack of variability in responses—were removed to enhance data reliability. The correlation matrix was used to assess multicollinearity and identify variables for exclusion or combination. A correlation coefficient above 0.7 typically indicates a high degree of multicollinearity, requiring further evaluation \cite{hair2013multivariate}.

Multivariate normality assesses the distribution of multiple variables at the same time. Multivariate outliers are data points that significantly deviate from the collective pattern, potentially distorting parameter estimates and introducing bias in predictive models. Unlike univariate outliers, which affect individual variables, multivariate outliers exhibit unusual combinations and relationships among multiple variables simultaneously. To detect such outliers, Mahalanobis distance and Cook’s distance were employed. The maximum Mahalanobis distance for the composite scores of C1, C2, C3, C4, and C5 was 20.557, exceeding the chi-square threshold of 9.488 at four degrees of freedom (95\% CI), indicating the presence of multivariate outliers, these datapoints were subsequently removed. Cook’s distance values were all below 1, so no observations were excluded based on this criterion. Additionally, Variance Inflation Factor (VIF) statistics for all constructs remained below 3, and collinearity tolerance values for all constructs were less than 1, confirming the absence of multicollinearity.

For the purpose of hypothesis testing, we developed eight hypotheses with five latent variables, consisting of eighteen observed variables on a five-point Likert scale. The model construct is depicted in Figure \ref{SEM modelZ}. One can wonder why the measuring model did not use every statement from the questionnaire included in the appendix. It is well known that in hypothesis testing, several statements in the questionnaire are not included in the analysis due to their failure to meet the required criteria (discriminant validity, CR and AVE conditions, factor loading of less than 0.60, or problematic internal linkages between items). This scenario may arise more frequently, particularly in online surveys. A small number of the study's questions were eliminated from analysis due to responses that did not fit the Likert scale. A few more also failed to meet the measurement model's validity and reliability requirements. 

\section{Data Analysis and Results} \label{sec: Result}
\subsection{Impact of Sociodemographic Features on Willingness to Use AI-enabled E-scooter}
\begin{figure}[!ht]
    \centering
    {\includegraphics[width=1\textwidth]
    {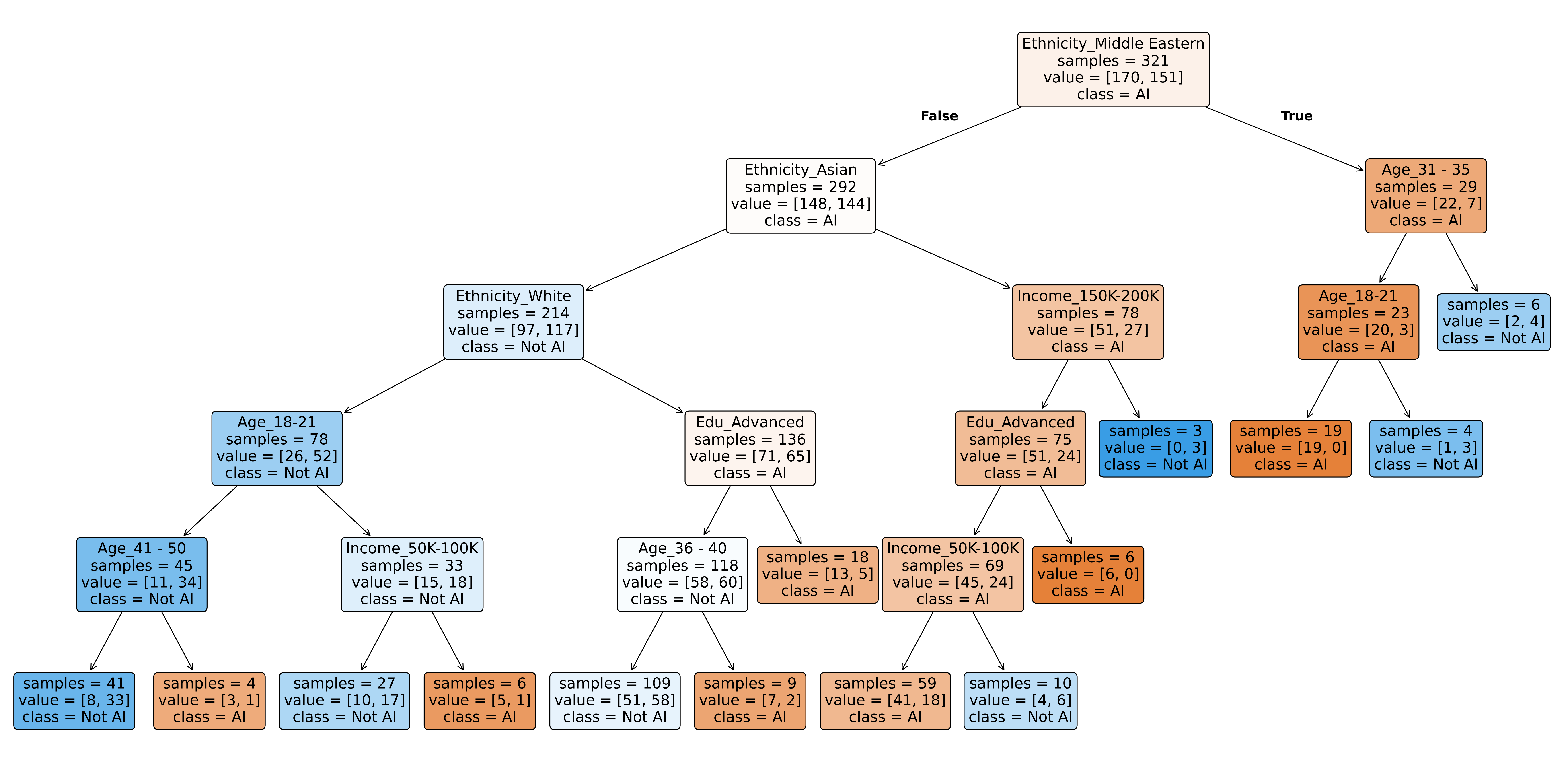}}
    \caption{Classification tree for demographic analysis of AI-assisted vs regular e-scooter choice}
    \label{fig:DecisionTree}
\end{figure}

Figure \ref{fig:DecisionTree} represents the final result of the classification tree. Each node in the decision tree contains several key elements: the "samples" value represents the number of data points reaching that node, the "value" shows the distribution of classes at the node (AI-assisted vs. Not AI-assisted), and the "class" indicates the predicted class for that node. 

The decision tree classification results reveal that ethnicity is the most influential feature in participants' preferences between AI-assisted and regular e-scooters. At the root node, participants of Middle Eastern ethnicity demonstrated a strong preference for AI-assisted e-scooters. However, within this group, individuals aged 31–35 and 18–21 showed a greater inclination toward regular options.

Following Middle Eastern participants, a significant proportion of Asian participants also leaned toward AI-assisted e-scooters, especially those holding advanced degrees such as doctorates or professional healthcare qualifications. Interestingly, within the Asian group, participants with higher income ranges (\$150k–\$200k) or even mid-range incomes (\$50k–\$100k) exhibited a tendency to prefer regular options.

For the White/Caucasian group, no strong overall preference was observed. However, education level emerged as an important factor, with individuals holding advanced degrees predominantly selecting AI-assisted e-scooters. Additionally, among White/Caucasian participants, those aged 36–40 were more likely to choose AI-assisted options.

Among other ethnic groups, including African American, Hispanic/ Latino, and others/ undisclosed, a larger proportion of participants generally preferred regular e-scooters. Within these groups, age emerged as a significant factor influencing their choices.
Younger participants aged 18–21 with lower incomes (less than \$50k) tended to favor non-AI-assisted e-scooters. However, within the same age range, those with incomes between \$50k - \$100k were more inclined to select AI-assisted e-scooters. Additionally, participants aged 41–50 predominantly opted for AI-assisted e-scooters, whereas those in other age ranges primarily preferred non-AI-assisted options.
In the following, we outline the conducted SEM analysis to validate the process and ensure that the constructs and their indicators meet the necessary reliability and validity criteria before hypothesis testing.

\subsection{Validation of SEM Measurement Model}
The analysis in Table \ref{tab:reliability-validity} indicates that the construct reliability (CR) and Cronbach's alpha (CA) are above 0.50 for each factor with and exception of C1  with CA=0.45 and CR=0.46 which is acceptable in this case, given the context. In their paper, An applied Orientation, Malhotra et al. \cite{malhotra2010applied} stated that average variance extracted (AVE) is often too strict, and reliability can be established through CR alone. The factor loading for each unobserved variable is 0.3 or above which is within the acceptable limit \citep{hair2019use, field2024discovering}.  Also, the AVE for C1 and C2 are 0.23 and 0.33 while for C3, C4 and C5 are 0.32, 0.35, and 0.60 respectively. The CR was found to be more than 0.6 for all the latent variables with the exception of C1 for which CR was calculated as 0.46.

\begin{table}[!ht]
\centering
\caption{Evaluation of Reliability and Validity of Latent Constructs}
\label{tab:reliability-validity}
\renewcommand{\arraystretch}{1.3} 
\setlength{\tabcolsep}{4pt} 
\resizebox{0.98\textwidth}{!}{ 
\begin{tabular}{p{4.5cm} c p{7.5cm} c c c c c}
\toprule
\textbf{Constructs} & \textbf{Code} & \textbf{Survey Questions} & \textbf{Mean} & \textbf{Factor Loading} & \textbf{CA} & \textbf{CR} & \textbf{AVE} \\ 
\midrule
\multirow{3}{=}{\raggedright Frequency of Regular Micromobility Use} 
& A & \raggedright Q1: How frequently do you use any of the following vehicles? - E-scooter & 1.64 & 0.47 & 0.45 & 0.46 & 0.23 \\  
& B & \raggedright Q2: How frequently do you use any of the following vehicles? - Bicycle & 1.92 & 0.35 &  &  &  \\  
& C & \raggedright Q3: How frequently do you use any of the following vehicles? - E-bike & 1.41 & 0.58 &  &  &  \\  
\midrule
\multirow{5}{=}{\raggedright Perceived Safety of Road Users Around E-Scooters} 
& F & \raggedright Q4: As a pedestrian, how safe do you feel when e-scooters are around you? & 3.38 & 0.53 & 0.72 & 0.71 & 0.33 \\  
& G & \raggedright Q5: When riding a bicycle, how safe do you feel when e-scooters are around you? & 3.66 & 0.65 &  &  &  \\  
& H & \raggedright Q6: When riding an e-bike, how safe do you feel when e-scooters are around you? & 3.91 & 0.61 &  &  &  \\  
& I & \raggedright Q7: When driving a car, how safe do you feel when e-scooters are around you? & 3.12 & 0.45 &  &  &  \\  
& J & \raggedright Q8: When riding an e-scooter, how safe do you feel when e-scooters are around you? & 3.79 & 0.61 &  &  &  \\  
\midrule
\multirow{3}{=}{\raggedright Perception of Safety in AI-Enabled Technology} 
& S & \raggedright Q9: How do you feel about the safety of these driving assistance technologies? & 3.47 & 0.45 & 0.6 & 0.61 & 0.32 \\  
& T & \raggedright Q10: Do you think autonomous vehicles (self-driving cars) offer a higher or safer level of safety compared to vehicles driven manually by humans? & 3.02 & 0.42 &  &  &  \\  
& W & \raggedright Q11: Compared to regular e-scooters, the AI-assisted features will reduce the likelihood of accidents. & 3.48 & 0.77 &  &  &  \\  
\midrule
\multirow{3}{=}{\raggedright Perception of Trust in AI-Enabled E-Scooters} 
& U & \raggedright Q12: I trust this system for a safer ride. & 2.97 & 0.78 & 0.78 & 0.78 & 0.55 \\  
& X & \raggedright Q13: I trust the AI's ability to handle unexpected situations while I'm on the e-scooter. & 3.11 & 0.78 &  &  &  \\  
& Y & \raggedright Q14: I trust that the AI won't malfunction and compromise my safety while I'm using the e-scooter. & 3.00 & 0.64 &  &  &  \\  
\midrule
\multirow{2}{=}{\raggedright Willingness to Use AI-Enabled E-Scooters} 
& V & \raggedright Q15: Compared to regular e-scooters, I feel more confident using the AI-assisted e-scooter in various traffic conditions because of its AI capabilities. & 3.41 & 0.84 & 0.7 & 0.74 & 0.60 \\  
& Z & \raggedright Q16: Given a choice between two e-scooters priced the same, which one would you buy/select from the deck? & 3.6 & 0.69 &  &  &  \\  
\bottomrule
\end{tabular}}
\end{table}

Table \ref{tab:constructs} represents the correlation matrix and discriminant validity for five constructs labeled as C1, C2, C3, C4, and C5. Each diagonal element (in bold) represents the square root of the average variance extracted (AVE) for the respective construct. The off-diagonal values represent the inter-construct correlations \cite{henseler2015new}. C1 and C2 demonstrate strong discriminant validity and appear to be well differentiated, while C3, C4, and C5 exhibit inter-construct correlations due to their conceptual overlap, as they represent closely related dimensions but on investigating the underlying conceptual definitions of these constructs, we can see that they are genuinely distinct categories and are conceptually valid. 

The correlation analysis reveals a strong positive relationship between trust and safety (0.738) and an even stronger correlation between trust and use (0.952), indicating that higher trust levels are closely linked to both perceived safety and willingness to use the system. Similarly, safety and use (1.037) show an unexpectedly high correlation, suggesting potential multicollinearity or measurement issues. Perspective and safety (0.143) exhibit a weak positive correlation, implying a minimal connection between user perspective and safety perception. Conversely, frequency of use and perceived safety of road users (-0.134) show a slight negative correlation, indicating that increased frequency of micromobility use may not necessarily enhance user perspectives on safety around e-scooters.
HTMT ratio analysis was also conducted to confirm discriminant validity using the HTMT $<$ 0.85 rule, as depicted in following section.

\begin{table}[!ht]
    \centering
    \footnotesize 
    \setlength{\tabcolsep}{3pt} 
    \renewcommand{\arraystretch}{1.2} 
    \caption{Correlation matrix and discriminant validity}
    \label{tab:constructs}
    \begin{tabular}{clccccc}
        \toprule
        \textbf{} & \textbf{Construct} & \textbf{C1} & \textbf{C2} & \textbf{C3} & \textbf{C4} & \textbf{C5} \\
        \midrule
        \textbf{1} & Frequency of regular micromobility use & \textbf{0.475} \\
        \textbf{2} & Perceived Safety of Road Users around E-Scooters & -0.134 & \textbf{0.571} \\
        \textbf{3} & Perception of Safety in AI-enabled technology & -0.024 & 0.143 & \textbf{0.566} \\
        \textbf{4} & Perception of Trust in AI-enabled E-Scooter & -0.031 & 0.176 & 0.982 & \textbf{0.738} \\
        \textbf{5} & Willingness to Use AI-enabled e-scooter & -0.089 & 0.063 & 0.967 & 0.952 & \textbf{0.768} \\
        \bottomrule
    \end{tabular}
    \captionsetup{justification=centering, font=footnotesize} 
    \caption*{\textit{Note: Bold numbers on the diagonal are square roots of AVE's.}}
\end{table}

The Hetrotrait-Monotrait Ratio (HTMT) matrix has been shown in Table \ref{tab:HMT}. It demonstrates the discriminant validity of the constructs (C1, C2, C3, C4, and C5). Based on the thresholds of 0.850 (strict) and 0.900 (liberal) \cite{fornell1981evaluating}, the majority of the construct pairs exhibit satisfactory discriminant validity. Specifically, the HTMT values between C1, C2, and other constructs are well below both thresholds, indicating strong evidence of discriminant validity. However, C3 and C4 (HTMT = 0.71) and C4 and C5 (HTMT = 0.707) are close to the liberal threshold but remain within acceptable limits. This suggests moderate overlap but still supports discriminant validity under both strict and liberal criteria. Overall, the matrix highlights that the constructs are distinct, with no violations of the thresholds, affirming the robustness of the measurement model.

\begin{table}[!ht]
    \centering
    \footnotesize
    \setlength{\tabcolsep}{3pt} 
    \renewcommand{\arraystretch}{1.15} 
    \caption{Heterotrait-Monotrait Ratio (HTMT)}
    \label{tab:HMT}
    \begin{tabular}{clccccc}
        \toprule
        \textbf{} & \textbf{Construct} & \textbf{C1} & \textbf{C2} & \textbf{C3} & \textbf{C4} & \textbf{C5} \\
        \midrule
        \textbf{1} & C1 & \textbf{-} \\
        \textbf{2} & C2 & 0.058 & \textbf{-} \\
        \textbf{3} & C3 & 0.001 & 0.120 & \textbf{-} \\
        \textbf{4} & C4 & 0.039 & 0.130 & 0.711 & \textbf{-} \\
        \textbf{5} & C5 & 0.081 & 0.003 & 0.655 & 0.707 & \textbf{-} \\
        \bottomrule
    \end{tabular}
    \captionsetup{justification=centering, font=footnotesize} 
    \caption*{\textit{Note:} Thresholds are 0.850 for strict and 0.900 for liberal discriminant validity.}
\end{table}

\begin{figure}[!ht]
    \centering
    \includegraphics[width=0.77\linewidth]{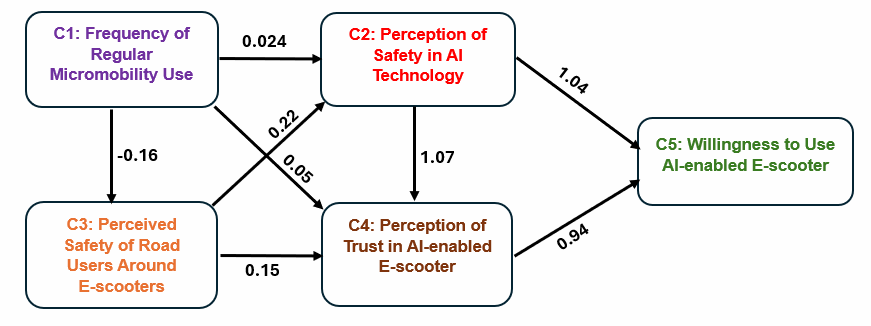}
    \caption{The proposed model to assess willingness to use AI enabled e-scooter}
    \label{SEM model2}
\end{figure}

\begin{table}[!ht]
\centering
\caption{Hypothesis Testing Results}
\label{tab:hypothesis-results}
\renewcommand{\arraystretch}{1.5} 
\setlength{\tabcolsep}{5pt} 
\footnotesize 
\resizebox{0.99\textwidth}{!}{ 
\begin{tabular}{p{7.5cm} c c c c}
\toprule
\textbf{Hypothesis} & \textbf{Label} & \textbf{Path Coefficient} ($\boldsymbol{\beta}$) & \textbf{p-value} & \textbf{Conclusion} \\ 
\midrule
\raggedright H1: Frequency of regular micromobility use influences road users' perception of safety in AI-enabled technology. 
& H1 & 0.024 & 0.798 & Not Supported \\ 

\raggedright H2: Frequency of regular micromobility use influences perceived safety of road users around e-scooters. 
& H2 & -0.155 & 0.096 & Not Supported \\ 

\raggedright H3: Frequency of regular micromobility use influences perception of trust in AI-enabled e-scooters. 
& H3 & -0.051 & 0.519 & Not Supported \\ 

\raggedright H4: Perceived safety of road users around e-scooters influences perception of safety in AI-enabled technology. 
& H4 & 0.218 & 0.009 & \textbf{Supported} \\ 

\raggedright H5: Perceived safety of road users around e-scooters influences perception of trust in AI-enabled e-scooters. 
& H5 & 0.145 & 0.033 & \textbf{Supported} \\ 

\raggedright H6: Perception of safety in AI-enabled technology influences perception of trust in AI-enabled e-scooters. 
& H6 & 1.075 & 0.001 & \textbf{Supported} \\ 

\raggedright H7: Perception of safety in AI-enabled technology influences willingness to use AI-enabled e-scooters. 
& H7 & 1.040 & 0.001 & \textbf{Supported} \\ 

\raggedright H8: Perception of trust in AI-enabled e-scooters influences willingness to use AI-enabled e-scooters. 
& H8 & 0.935 & 0.001 & \textbf{Supported} \\ 
\bottomrule
\end{tabular}}
\end{table}

\subsection{Hypothesis Testing Results from Structural Equation Modeling}
This section presents the results of the Structural Equation Modeling (SEM) analysis, focusing on the direct effects of the hypothesized relationships. The standardized regression weights (SRW) and p-values indicate the strength and statistical significance of each relationship. The results are shown in Table \ref{tab:hypothesis-results} and interpreted as follows:

The first hypothesis (H1) examined the relationship between frequency of regular micromobility use (C1) and perception of safety in AI-enabled technology (C3). The standardized estimate of 0.024 suggests a weak positive effect; however, the high p-value of 0.798 indicates that this relationship is not statistically significant. Consequently, there is no strong evidence to suggest that frequency of micromobility use significantly influences perceptions of AI safety. Similarly, H2 tested the impact of C1 on perceived safety of road users around e-scooters (C2). The results indicate a weak negative relationship (SRW = -0.134, p = 0.096), which does not reach statistical significance at the 95\% confidence level. Although this near-threshold p-value suggests a potential trend, the effect remains small and may require a larger sample size for conclusive results.

For H3, the relationship between C1 and perception of trust in AI-enabled e-scooters (C4) was analyzed. The unstandardized coefficient of -0.063 and a standardized regression weight of -0.051 indicate an extremely weak negative relationship, further supported by a non-significant p-value of 0.519. This suggests that frequency of micromobility use does not have a meaningful impact on trust in AI-enabled e-scooters. In contrast, H4 revealed a moderate positive relationship between perceived safety of road users (C2) and perception of safety in AI-enabled technology (C3). The standardized regression weight of 0.218 with a p-value of 0.009 confirms a statistically significant effect, suggesting that individuals who perceive greater safety around e-scooters are also more likely to perceive AI-assisted technologies as safe.

H5 examined the influence of C2 on C4, indicating a weak but statistically significant relationship (SRW = 0.145, p = 0.033). This result suggests that perceived safety of road users has a small but noticeable positive influence on trust in AI-enabled e-scooters. Although the effect size is not substantial, the statistical significance confirms that perceptions of general road safety contribute to AI trust formation.

A particularly strong relationship was observed for H6, which tested the effect of perception of safety in AI-enabled technology (C3) on perception of trust in AI-enabled e-scooters (C4). The standardized regression weight of 1.07 suggests a substantial positive effect, and the relationship is highly statistically significant \(p < 0.001\). These findings indicate that perceptions of AI safety strongly enhance user trust in AI-enabled e-scooters, making C3 a critical predictor of C4.

The results for H7 further reinforce the importance of AI safety perceptions in user adoption. A standardized regression weight of 1.04 \(p < 0.001\) indicates a strong and significant positive relationship between C3 and willingness to use AI-enabled e-scooters (C5). This suggests that as confidence in AI-assisted micromobility safety increases, users are substantially more willing to adopt such technology. Similarly, H8 confirmed a strong positive effect between C4 and C5, with a standardized regression weight of 0.935 \(p < 0.001\). This result suggests that higher trust in AI-enabled e-scooters significantly increases willingness to use them.

Overall, the hypothesis testing results reveal that frequency of micromobility use (C1) does not significantly impact safety perceptions or trust in AI-assisted e-scooters, as evidenced by the non-significant results in H1–H3. However, perceived safety of road users (C2) positively influences both AI safety perceptions (C3) and trust in AI-enabled e-scooters (C4), as seen in H4 and H5. The most influential factors driving willingness to use AI-enabled e-scooters (C5) are perceptions of AI safety (C3) and trust in AI-enabled e-scooters (C4), which demonstrated strong and highly significant relationships in H6–H8. These findings suggest that enhancing user trust and AI safety perceptions may be more effective in increasing AI micromobility adoption than simply relying on prior micromobility usage experience. Future research should explore potential moderating factors that may refine these relationships further.

\subsection{Full Scale Structure Equation Modeling}
The five constructs and their item descriptions are illustrated in Table \ref{tab:regression_results}. We used Cronbach's Alpha (CA) scale to check the reliability of the full scale model, the summary statistics are - CA =0.759, mean = 3.088, with a minimum value of 1.418 and maximum value of 3.919, range = 2.501 with a variance of 0.573. The CR value for each construct C1, C2, C3, C4, and C5 in the model were calculated to be 0.48, 0.72, 0.73, 0.81, and 0.53 respectively. We also included an independent construct in the full model, C6: "Crash Experience" which consisted of two items -
\begin{itemize}
    \item Have you ever been involved in an accident while riding an e-scooter?
    \item How would you categorize the severity of your injury from the e-scooter incident?
\end{itemize}
The summary statistics for this construct consisted of Cronbach's alpha = 0.676, cumulative mean of 4.81, SD = 0.65, and variance =0.13. 

The full-scale model consists of five exogenous variables C1, C2, C3, C4, and C6 which have an effect on one endogenous variable i.e. variable C5. Since, Willingness to use AI enabled e-scooter. (C5) is a dependent variable, it needs to have an error term called disturbance added to it, i.e. e21 depicted in the full-scale figure. 

\begin{table}[h!]
\centering
\caption{Regression Results}
\resizebox{0.9\textwidth}{!}{
\begin{tabular}{lccccc}
\toprule
\textbf{Predictor} & \textbf{Standardized Estimate} & \textbf{S.E.} & \textbf{C.R.} & \textbf{p-value} & \textbf{Interpretation} \\ \midrule
C5 $\leftarrow$ C2 & -0.091 & 0.07 & -1.674 & 0.094 & Negative, marginally significant \\
C5 $\leftarrow$ C4 & 0.597 & 0.059 & 9.46 & 0.001 & Strong positive, highly significant \\
C5 $\leftarrow$ C3 & 0.775 & 0.147 & 6.176 & 0.001 & Strong positive, highly significant \\
C5 $\leftarrow$ C1 & -0.089 & 0.093 & -1.296 & 0.195 & Negative, non-significant \\
C5 $\leftarrow$ C6 & -0.621 & 12.039 & -0.336 & 0.737 & Negative, non-significant \\ \bottomrule
\end{tabular}
}

\label{tab:regression_results}
\end{table}

The summary of the key findings from the SEM results regarding the direct effects of the latent variables (C1, C2, C3, C4, and C6) on the dependent variable C5 are depicted below in figure \ref{SEM Direct Effects Full ModelZ}.

\begin{figure}[H]
    \centering
    \includegraphics[width=0.65\textwidth]{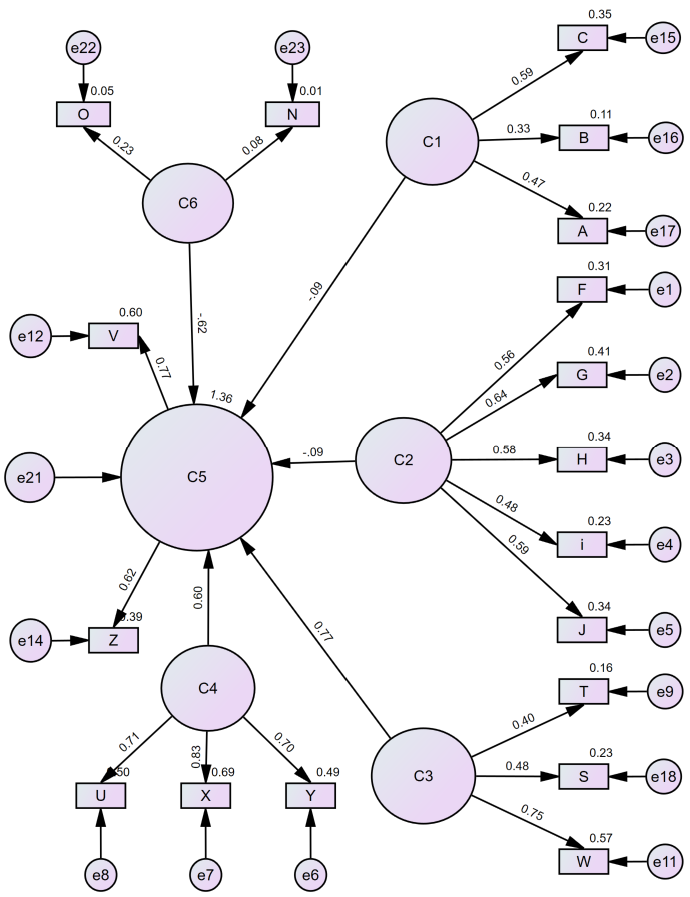} 
    \caption{Structure Equation Modeling Full Scale Model}
    \label{SEM Direct Effects Full ModelZ}
\end{figure}

Effect of C1 (frequency of regular micromobility use) on C5 (Willingness to use AI-enabled e-scooter): The standardized regression weight (–0.089) indicates a weak negative relationship between frequency of regular micromobility use and willingness to use AI-enabled e-scooter. The critical ratio (C.R. = -1.296) and p-value (0.195) indicate that this effect is not statistically significant. C1 does not have a meaningful impact on C5 in this model, and the relationship may not be reliable. It could be because people who have been using other modes of micromobility like e-scooters, bicycles, and e-bikes, may not be willing to use AI-enabled technology because they are already comfortable with their current mode of locomotion. 
Effect of C2 (perceived safety of road users around e-scooters) and C5 (willingness to use AI-enabled e-scooter): The standardized regression weight (–0.091) indicates a slight negative relationship between perceived safety of road users around e-scooters and willingness to use AI-enabled e-scooter. The critical ratio (C.R. = -1.674) and p-value (0.094) indicate that this effect is not statistically significant. C2 does not appear to have a meaningful impact on C5 in this model.
Effect of C3 (perception of safety in AI-enabled technology) on C5 (willingness to use AI-enabled e-scooter): The standardized regression weight (0.775) indicates a strong positive relationship between perception of safety in AI-enabled technology and willingness to use AI-enabled e-scooter. For every 1-unit increase in C3, C5 is expected to increase by 0.775 standard deviations in the standardized scale. The effect is highly significant (C.R. = 6.176, $p < 0.001$). This demonstrates that C3 is a major determinant of C5 with a substantial impact. 
Effect of C4 (perception of trust in AI-enabled E-Scooter) on C5 (willingness to use AI-enabled e-scooter): The standardized regression weight (0.597) indicates a positive relationship between perception of trust in AI-enabled E-Scooter and willingness to use AI-enabled e-scooter. For every 1-unit increase in C4, C5 is expected to increase by 0.597 standard deviations in the standardized scale. The effect is highly significant (C.R. = 9.46, $p < 0.001$). This suggests that C4 is an important and statistically significant predictor of C5.
Effect of C6 (Crash Experience) on C5 (willingness to use AI-enabled e-scooter): The standardized regression weight (–0.621) indicates a slight negative relationship between previous Crash experience and Willingness to Use AI-enabled e-scooter. This seems logical because people who have been previously involved in an accident using an E-scooter, may not be interested in using it again even with the AI-enabled features. The critical ratio (C.R. = -0.336) and p-value (0.621) suggest that this effect is not statistically significant. Crash experience does not significantly influence Willingness to use AI enabled e-scooter.

By observing the findings from the Full SEM model, we can see that perception of safety in AI-enabled technologies has the strongest positive effect on Willingness to use AI-enabled e-scooter, followed by perception of trust in AI-enabled E-Scooters, both of which are statistically significant. Both perception of safety and trust are significant predictors while frequency of regular micromobility use, perceived safety of road users around e-scooters, and Crash experience do not significantly affect willingness to use AI-enabled e-scooters. Their p-values are greater than 0.05, and their critical ratios are not substantial enough to indicate statistical significance. For the purpose of practical implications, efforts to influence willingness to use AI-enabled e-scooters should focus on factors that enhance perception of safety and perception of trust, as these are the most impactful predictors. Variables like frequeny of micromobility use, perceived safety of road users around e-scooters, and previous crash experience are not the driving factors in the adoption and willingness to use AI-enabled e-scooters.

\section{Discussion} \label{sec:discussion}
\subsection{Impact of Sociodemographic Factors on Adoption of AI-Assisted E-Scooters}
Sociodemographic analysis revealed that gender was not an influential factor in participants' inclination towards AI-assisted e-scooters. This finding contrasts with previous studies indicating that males are more likely to use e-scooters, a trend linked to the gender mobility gap and males' greater risk-taking behavior \cite{laa2020survey,nikiforiadis2021analysis}. However, AI-assisted e-scooters could potentially mitigate risks, serving as a tool to narrow the gender disparity in e-scooter adoption.
Younger adults, especially those aged 18-21, showed a stronger preference for regular e-scooters. 
This aligns with research suggesting that older adults perceive more benefits from AI-assisted vehicles than younger individuals \cite{hilgarter2020public}.
The tendency of younger riders to avoid AI-assisted e-scooters could be attributed to their greater propensity for risky behaviors while riding, as highlighted by \cite{gioldasis2021risk}. Additionally, e-scooter trauma patients tend to be younger \cite{burt2023scooter}, reinforcing the notion that younger riders may exhibit more risk-taking behavior, which AI-assisted technology could help mitigate.

Education also played a significant role in individuals' preferences, particularly among those with advanced degrees, including doctorate degrees (e.g., PhD, EdD) or professional healthcare degrees (e.g., MD, DDS, DVM). This preference could be attributed to their greater exposure to AI knowledge and its applications. Individuals with such educational backgrounds are often more engaged in research and are regularly exposed to the benefits and advancements of AI across various scientific fields, making them more likely to appreciate the potential advantages of AI-assisted e-scooters. 

\subsection{Influence of Micromobility Use and Crash Experience on AI-Assisted E-Scooter Preferences}
Findings from the hypothesis testing revealed that H1 which proposed that frequent use of micormobility vehicles positively influences perceptions of safety in AI-enabled technologies, was not supported.
A Previous study by Pourfalatoun et al. \cite{pourfalatoun2023shared} has shown that e-scooter users were more likely to be early adopters of new technology. While the analysis yielded a non-significant relationship ($\beta$ = 0.024, p = 0.798), the positive coefficient suggests a potential association, the lack of statistical significance implies that familiarity with micromobility might not inherently translate to a more favorable attitude towards AI safety technology. This could be due to variations in user experiences with micromobility and AI technologies or a general lack of public awareness connecting the two domains. Similarly, H2 examined whether frequent micromobility use affects perceived safety of road users around e-scooters. The negative coefficient ($\beta$ = -0.155, p = 0.096) aligns with the previous study \cite{nikiforiadis2021analysis} and our hypothesis that it does have an effect which is negative in nature, but the lack of statistical significance again points to a non-conclusive relationship. This weak relationship could stem from diverse factors such as variations in e-scooter regulations, cultural norms, or differences in micromobility vehicle design that dilute a clear pattern of perception. This implies that frequent users' experiences with micromobiliy do not strongly shape their broader opinion about E-scooter use. 

In the full model, it seems logical that crash experience had a negative effect on willingness to use AI-enabled e-scooter. This finding is in line with previous studies \cite{he2017impact} which found prior accidents had a significant impact on the preference of travel mode where the riders were reported to have greater fears for personal safety, worries about driving, and negative physical and physiological symptoms \cite{marasini2022psychological}. Individuals who have been involved in accidents while riding conventional e-scooters may develop a reluctance to continue using such vehicles, even if AI-assisted features are introduced. This finding underscores the potential psychological barriers to AI adoption among riders with prior negative experiences. 

Similarly, perceived safety of road users around e-scooters negatively influenced willingness to use AI-enabled e-scooters. This suggests that general attitudes toward regular e-scooters, whether positive or negative, do not necessarily extend to AI-assisted versions. Users may perceive AI-enabled e-scooters as fundamentally different from traditional ones, making their pre-existing perceptions of conventional e-scooters less relevant to their willingness to adopt AI-based alternatives.

Finally, the frequency of other types of micromobility use (e-scooters, bicycles, and e-bikes) was negatively associated with willingness to adopt AI-assisted e-scooters. This is in line with similar findings by Sharpe \cite{sharpe2013aesthetics} who showed that people may continue using familiar transport modes due to habitual reinforcement. This indicates that users who are already accustomed to their current micromobility vehicles may be less inclined to transition to AI-assisted models. Resistance to change \cite{lattarulo2019resistance}, comfort with familiar transportation modes, and skepticism about AI enhancements could contribute to this reluctance, even when AI features offer potential safety improvements.

\subsection{Implications for Urban Mobility and AI Adoption}

The findings of this study highlight several important considerations for urban mobility stakeholders, policymakers, and AI developers. Firstly, AI-assisted e-scooters may serve as a tool to reduce existing demographic disparities in micromobility adoption, particularly regarding gender and safety concerns. By enhancing safety features, AI-assisted e-scooters may appeal more to females and other underrepresented groups, contributing to a more equitable urban mobility landscape.
Secondly, educational initiatives aimed at improving AI literacy could enhance public trust and acceptance of AI-driven transportation solutions. Increasing awareness about the benefits and capabilities of AI technology can help mitigate skepticism and foster a more receptive environment for AI-assisted e-scooters.
Thirdly, strategies to overcome psychological barriers among users with prior crash experiences should be explored, as safety-related trauma may hinder the adoption of AI-assisted e-scooters, underscoring the need for targeted interventions to address these concerns. By addressing these challenges, stakeholders can develop more effective strategies to promote the adoption of AI-assisted e-scooters and enhance urban mobility.


\section{Conclusion} \label{sec:conclusion}
This study examined the factors influencing the adoption of AI-assisted e-scooters, focusing on sociodemographics, safety perceptions, trust, and prior micromobility experiences. The result of decision tree analysis on demographics indicates that ethnicity, income, and age, significantly shape adoption preferences, with Middle Eastern and Asian participants showing greater willingness to use AI-enabled e-scooters, while younger, lower-income individuals from other ethnic groups tended to prefer regular e-scooters.
Results from the hypothesis testing revealed that road users’ perceptions of e-scooters demonstrated meaningful associations with safety and trust. Perceived safety of road users around e-scooters was linked to higher perceived safety in AI-enabled technologies and greater trust in AI-enabled e-scooters. These findings highlight the role of positive road users' interaction with regular e-scooters in shaping AI adoption. Conversely, the frequency of regular micromobility use did not significantly impact perceptions of safety in AI-enabled technologies, e-scooters, or trust in AI-enabled e-scooters.

It should be noted that perception of safety in AI-enabled technologies strongly influences both trust in AI-enabled e-scooters and willingness to use them. Moreover, trust in AI-enabled e-scooters significantly affects willingness to use them. These results highlight the foundational role of safety and trust in AI-assisted e-scooter adoption.


Perceived safety in AI-enabled technology emerged as the strongest predictor of willingness to use AI-assisted e-scooters closely followed by perception of trust in AI-enabled e-scooter which showed a highly significant and strong positive relationship with willingness to use AI-enabled e-scooter. In contrast, perceived safety of road users around e-scooters, frequency of regular micromobility use, and previous crash experience showed weaker negative associations that were statistically insignificant. This reinforces the dominant role of safety and trust perceptions over other factors in influencing AI-assisted e-scooter adoption.

From a policy and industry perspective, fostering trust in AI-assisted micromobility requires more than just increased exposure. Rather than assuming that frequent micromobility users will naturally transition to AI-assisted models, stakeholders should focus on enhancing safety perceptions and transparency in AI functionality. Given the strong influence of perceived safety on both trust and adoption, manufacturers and urban mobility planners must prioritize clear communication of AI safety features, real-world performance data, and user education programs.

\section{Limitation and Future Work} \label{sec:limitation}

The SEM direct effects models typically assume unidirectional relationships and are not designed to accommodate feedback loops, limiting their ability to capture the reciprocal causality often present in real-world systems. 
Contextual or situational factors, such as time or environment, that might moderate relationships between variables were not accommodated in this study. Furthermore, SEM models are highly context-dependent, meaning that the findings from a specific model may not generalize to other populations or settings without further validation. 
Furthermore, future studies should include a more diverse group of respondents in terms of age and ethnicity to develop a deeper understanding of public perception regarding these technologies. There is a significant gap in the literature concerning adolescents' perception of safety and trust in using e-scooters, particularly as this mode of transport gains popularity among this age group, coinciding with the rise in e-scooter-related injuries among adolescents \cite{douglas2024high}.
Future research could explore the development of a high-fidelity virtual reality (VR) simulator of an AI-enabled e-scooter, allowing participants to virtually experience how the system functions \cite{guo2023psycho, MAD-IVE}. The 3D simulation in VR can enable users to experience an almost real-life scenario in a safe, environment, allowing the researchers to access the impact of AI-enabled e-scooters on users' perceptions of safety, trust, and usability. Furtehrmore, recent research has demonstrated the potential of multimodal augmented reality safety warnings in improving situational awareness and reducing reaction times in workers \cite{SABETI2024104867, Sabeti02012024} AR’s capability in enhancing real-time safety interventions. Similarly, EEG analysis was used to measure cognitive responses to AR safety warnings \cite{BANANIARDECANI2025106802}, revealing improved attention and situational awareness in high-risk environments. Future research on AI-assisted e-scooters could leverage these findings by integrating multi modal safety alerts into e-scooter systems, and study the human factors of warnings in the context of micromobiliy. Such an approach could be used to evaluate warnings impact on users' trust and perception of safety in AI-assisted micromobility systems.


 \bibliographystyle{elsarticle-num} 
 \bibliography{cas-refs}

\begin{thebibliography}{10}
\expandafter\ifx\csname url\endcsname\relax
  \def\url#1{\texttt{#1}}\fi
\expandafter\ifx\csname urlprefix\endcsname\relax\def\urlprefix{URL }\fi
\expandafter\ifx\csname href\endcsname\relax
  \def\href#1#2{#2} \def\path#1{#1}\fi

\bibitem{shaheen2020sharing}
S.~Shaheen, A.~Cohen, N.~Chan, A.~Bansal, Sharing strategies: carsharing, shared micromobility (bikesharing and scooter sharing), transportation network companies, microtransit, and other innovative mobility modes, in: Transportation, land use, and environmental planning, Elsevier, 2020, pp. 237--262.

\bibitem{nacto2024shared}
J.~NACTO, A micromobility record: 157 million trips on bike share and scooter share in 2023, National Association of City Transportation Officials (2024).

\bibitem{burt2023scooter}
N.~Burt, Z.~Ahmed, E-scooter attitudes and risk-taking behaviours: an international systematic literature review and survey responses in the west midlands, united kingdom, Frontiers in public health 11 (2023) 1277378.

\bibitem{hardt2019usage}
C.~Hardt, K.~Bogenberger, Usage of e-scooters in urban environments, Transportation research procedia 37 (2019) 155--162.

\bibitem{comi2022innovative}
A.~Comi, A.~Polimeni, A.~Nuzzolo, An innovative methodology for micro-mobility network planning, Transportation research procedia 60 (2022) 20--27.

\bibitem{chen2024impact}
D.~Chen, A.~Hosseini, A.~Smith, D.~Xiang, A.~Heydarian, O.~Shoghli, B.~Campbell, Impact of different infrastructures and traffic scenarios on behavioral and physiological responses of e-scooter users, arXiv preprint arXiv:2407.10310 (2024).

\bibitem{sievert2023survey}
K.~Sievert, M.~Roen, C.~M. Craig, N.~L. Morris, A survey of electric-scooter riders’ route choice, safety perception, and helmet use, Sustainability 15~(8) (2023) 6609.

\bibitem{arun2021systematic}
A.~Arun, M.~M. Haque, S.~Washington, T.~Sayed, F.~Mannering, A systematic review of traffic conflict-based safety measures with a focus on application context, Analytic methods in accident research 32 (2021) 100185.

\bibitem{trivedi2019craniofacial}
B.~Trivedi, M.~J. Kesterke, R.~Bhattacharjee, W.~Weber, K.~Mynar, L.~V. Reddy, Craniofacial injuries seen with the introduction of bicycle-share electric scooters in an urban setting, Journal of Oral and Maxillofacial Surgery 77~(11) (2019) 2292--2297.

\bibitem{serra2021head}
G.~F. Serra, F.~A. Fernandes, E.~Noronha, R.~J.~A. de~Sousa, Head protection in electric micromobility: A critical review, recommendations, and future trends, Accident Analysis \& Prevention 163 (2021) 106430.

\bibitem{kazemzadeh2023electric}
K.~Kazemzadeh, M.~Haghani, F.~Sprei, Electric scooter safety: An integrative review of evidence from transport and medical research domains, Sustainable Cities and Society 89 (2023) 104313.

\bibitem{haworth2021comparing}
N.~Haworth, A.~Schramm, D.~Twisk, Comparing the risky behaviours of shared and private e-scooter and bicycle riders in downtown brisbane, australia, Accident Analysis \& Prevention 152 (2021) 105981.

\bibitem{feng2022estimating}
C.~Feng, J.~Jiao, H.~Wang, Estimating e-scooter traffic flow using big data to support planning for micromobility, Journal of Urban Technology 29~(2) (2022) 139--157.

\bibitem{greenblatt2015automated}
J.~B. Greenblatt, S.~Shaheen, Automated vehicles, on-demand mobility, and environmental impacts, Current sustainable/renewable energy reports 2 (2015) 74--81.

\bibitem{chen2024performance}
D.~Chen, A.~Hosseini, A.~Smith, A.~F. Nikkhah, A.~Heydarian, O.~Shoghli, B.~Campbell, Performance evaluation of real-time object detection for electric scooters, arXiv preprint arXiv:2405.03039 (2024).

\bibitem{li2023modeling}
T.~Li, J.~Kovaceva, M.~Dozza, Modeling collision avoidance maneuvers for micromobility vehicles, Journal of safety research 87 (2023) 232--243.

\bibitem{mckenzie2020urban}
G.~McKenzie, Urban mobility in the sharing economy: A spatiotemporal comparison of shared mobility services, Computers, Environment and Urban Systems 79 (2020) 101418.

\bibitem{rejali2021assessing}
S.~Rejali, K.~Aghabayk, A.~Mohammadi, N.~Shiwakoti, Assessing a priori acceptance of shared dockless e-scooters in iran, Transportation Research Part D: Transport and Environment 100 (2021) 103042.

\bibitem{samadzad2023factors}
M.~Samadzad, H.~Nosratzadeh, H.~Karami, A.~Karami, What are the factors affecting the adoption and use of electric scooter sharing systems from the end user's perspective?, Transport policy 136 (2023) 70--82.

\bibitem{ari2024investigating}
E.~Ar{\i}, V.~Y{\i}lmaz, Investigating the factors affecting electric scooter usage behavior with a proposed structural model, Research in Transportation Business \& Management 56 (2024) 101164.

\bibitem{pourfalatoun2023shared}
S.~Pourfalatoun, J.~Ahmed, E.~E. Miller, Shared electric scooter users and non-users: Perceptions on safety, adoption and risk, Sustainability 15~(11) (2023) 9045.

\bibitem{tian2022characteristics}
D.~Tian, A.~D. Ryan, C.~M. Craig, K.~Sievert, N.~L. Morris, Characteristics and risk factors for electric scooter-related crashes and injury crashes among scooter riders: A two-phase survey study, International journal of environmental research and public health 19~(16) (2022) 10129.

\bibitem{vsucha2023scooter}
M.~{\v{S}}ucha, E.~Drimlov{\'a}, K.~Re{\v{c}}ka, N.~Haworth, K.~Karlsen, A.~Fyhri, P.~Wallgren, P.~Silverans, F.~Slootmans, E-scooter riders and pedestrians: attitudes and interactions in five countries, Heliyon 9~(4) (2023).

\bibitem{derrick2020perceptions}
W.~T.~C. Derrick, Perceptions and attitudes towards micro e-scooters in singapore, Perceptions and attitudes towards micro e-scooters in Singapore (2020).

\bibitem{james2023pedestrians}
O.~James, J.~I. Swiderski, J.~Hicks, D.~Teoman, R.~Buehler, Pedestrians and e-scooters: An initial look at e-scooter parking and perceptions by riders and non-riders, Urban Affairs and Planning, Virginia TechCorrespondence: ralphbu@vt.edu (2023).

\bibitem{othman2021public}
K.~Othman, Public acceptance and perception of autonomous vehicles: a comprehensive review, AI and Ethics 1~(3) (2021) 355--387.

\bibitem{thomas2020perception}
E.~Thomas, C.~McCrudden, Z.~Wharton, A.~Behera, Perception of autonomous vehicles by the modern society: a survey, IET Intelligent Transport Systems 14~(10) (2020) 1228--1239.

\bibitem{pyrialakou2020perceptions}
V.~D. Pyrialakou, C.~Gkartzonikas, J.~D. Gatlin, K.~Gkritza, Perceptions of safety on a shared road: Driving, cycling, or walking near an autonomous vehicle, Journal of safety research 72 (2020) 249--258.

\bibitem{sanders2020scoot}
R.~L. Sanders, M.~Branion-Calles, T.~A. Nelson, To scoot or not to scoot: Findings from a recent survey about the benefits and barriers of using e-scooters for riders and non-riders, Transportation Research Part A: Policy and Practice 139 (2020) 217--227.

\bibitem{nikiforiadis2021analysis}
A.~Nikiforiadis, E.~Paschalidis, N.~Stamatiadis, A.~Raptopoulou, A.~Kostareli, S.~Basbas, Analysis of attitudes and engagement of shared e-scooter users, Transportation research part D: transport and environment 94 (2021) 102790.

\bibitem{teixeira2023barriers}
J.~F. Teixeira, V.~Diogo, A.~Bern{\'a}t, A.~Lukasiewicz, E.~Vaiciukynaite, V.~S. Sanna, Barriers to bike and e-scooter sharing usage: an analysis of non-users from five european capital cities, Case studies on transport policy 13 (2023) 101045.

\bibitem{she2017barriers}
Z.-Y. She, Q.~Sun, J.-J. Ma, B.-C. Xie, What are the barriers to widespread adoption of battery electric vehicles? a survey of public perception in tianjin, china, Transport Policy 56 (2017) 29--40.

\bibitem{javadinasr2022eliciting}
M.~Javadinasr, S.~Asgharpour, E.~Rahimi, P.~Choobchian, A.~K. Mohammadian, J.~Auld, Eliciting attitudinal factors affecting the continuance use of e-scooters: An empirical study in chicago, Transportation research part F: traffic psychology and behaviour 87 (2022) 87--101.

\bibitem{blazanin2022scooter}
G.~Blazanin, A.~Mondal, K.~E. Asmussen, C.~R. Bhat, E-scooter sharing and bikesharing systems: An individual-level analysis of factors affecting first-use and use frequency, Transportation research part C: emerging technologies 135 (2022) 103515.

\bibitem{bachand2012better}
J.~Bachand-Marleau, B.~H. Lee, A.~M. El-Geneidy, Better understanding of factors influencing likelihood of using shared bicycle systems and frequency of use, Transportation Research Record 2314~(1) (2012) 66--71.

\bibitem{xu2019analysis}
C.~Xu, C.~Wang, Analysis of e-bike trip duration and frequency by bayesian duration and zero-inflated count models, KSCE Journal of Civil Engineering 23 (2019) 1806--1818.

\bibitem{alp2025micromobility}
S.~Alp, Micromobility and artificial intelligence, in: Micromobility: Perspectives from Engineering, Urban Planning, Health Sciences and Social Sciences, Springer, 2025, pp. 315--328.

\bibitem{jevinger2024artificial}
{\AA}.~Jevinger, C.~Zhao, J.~A. Persson, P.~Davidsson, Artificial intelligence for improving public transport: a mapping study, Public Transport 16~(1) (2024) 99--158.

\bibitem{kuberkar2020factors}
S.~Kuberkar, T.~K. Singhal, Factors influencing adoption intention of ai powered chatbot for public transport services within a smart city, International Journal of Emerging Technologies in Learning 11~(3) (2020) 948--958.

\bibitem{azarbayjani2024trust}
A.~Azarbayjani, Trust and perceived safety of vulnerable adult road users towards regular and ai-enabled e-scooters, Master's thesis, The University of North Carolina at Charlotte (2024).

\bibitem{rousseau1998not}
D.~M. Rousseau, S.~B. Sitkin, R.~S. Burt, C.~Camerer, Not so different after all: A cross-discipline view of trust, Academy of management review 23~(3) (1998) 393--404.

\bibitem{hohenberger2017not}
C.~Hohenberger, M.~Sp{\"o}rrle, I.~M. Welpe, Not fearless, but self-enhanced: The effects of anxiety on the willingness to use autonomous cars depend on individual levels of self-enhancement, Technological Forecasting and Social Change 116 (2017) 40--52.

\bibitem{priyam2013comparative}
A.~Priyam, G.~R. Abhijeeta, A.~Rathee, S.~Srivastava, Comparative analysis of decision tree classification algorithms, International Journal of current engineering and technology 3~(2) (2013) 334--337.

\bibitem{de2013decision}
B.~De~Ville, Decision trees, Wiley Interdisciplinary Reviews: Computational Statistics 5~(6) (2013) 448--455.

\bibitem{bramer2007avoiding}
M.~Bramer, Avoiding overfitting of decision trees, Principles of data mining (2007) 119--134.

\bibitem{ullman2012structural}
J.~B. Ullman, P.~M. Bentler, Structural equation modeling, Handbook of Psychology, Second Edition 2 (2012).

\bibitem{kline2023principles}
R.~B. Kline, Principles and practice of structural equation modeling, Guilford publications, 2023.

\bibitem{SmartPLS4}
C.~M. Ringle, S.~Wende, J.-M. Becker, \href{{https://www.smartpls.com/}}{Smartpls 4} (2024).
\newline\urlprefix\url{{https://www.smartpls.com/}}

\bibitem{eisinga2013reliability}
R.~Eisinga, M.~t. Grotenhuis, B.~Pelzer, The reliability of a two-item scale: Pearson, cronbach, or spearman-brown?, International journal of public health 58 (2013) 637--642.

\bibitem{fornell1981evaluating}
C.~Fornell, D.~F. Larcker, Evaluating structural equation models with unobservable variables and measurement error, Journal of marketing research 18~(1) (1981) 39--50.

\bibitem{hair2019use}
J.~F. Hair, J.~J. Risher, M.~Sarstedt, C.~M. Ringle, When to use and how to report the results of pls-sem, European business review 31~(1) (2019) 2--24.

\bibitem{diba2023autonomous}
D.~S. Diba, N.~Gore, S.~S. Pulugurtha, et~al., Autonomous shuttle implementation and best practices [supporting dataset], Tech. rep., Mineta Transportation Institute (2023).

\bibitem{hair2013multivariate}
J.~F. Hair, W.~C. Black, B.~J. Babin, R.~E. Anderson, Multivariate data analysis: Pearson new international edition PDF eBook, Pearson Higher Ed, 2013.

\bibitem{malhotra2010applied}
N.~K. Malhotra, S.~Dash, An applied orientation, Marketing Research 2~(4) (2010) 109--122.

\bibitem{field2024discovering}
A.~Field, Discovering statistics using IBM SPSS statistics, Sage publications limited, 2024.

\bibitem{henseler2015new}
J.~Henseler, C.~M. Ringle, M.~Sarstedt, A new criterion for assessing discriminant validity in variance-based structural equation modeling, Journal of the academy of marketing science 43 (2015) 115--135.

\bibitem{laa2020survey}
B.~Laa, U.~Leth, Survey of e-scooter users in vienna: Who they are and how they ride, Journal of transport geography 89 (2020) 102874.

\bibitem{hilgarter2020public}
K.~Hilgarter, P.~Granig, Public perception of autonomous vehicles: A qualitative study based on interviews after riding an autonomous shuttle, Transportation research part F: traffic psychology and behaviour 72 (2020) 226--243.

\bibitem{gioldasis2021risk}
C.~Gioldasis, Z.~Christoforou, R.~Seidowsky, Risk-taking behaviors of e-scooter users: A survey in paris, Accident Analysis \& Prevention 163 (2021) 106427.

\bibitem{he2017impact}
S.~Y. He, J.~Th{\o}gersen, The impact of attitudes and perceptions on travel mode choice and car ownership in a chinese megacity: The case of guangzhou, Research in transportation economics 62 (2017) 57--67.

\bibitem{marasini2022psychological}
G.~Marasini, F.~Caleffi, L.~M. Machado, B.~M. Pereira, Psychological consequences of motor vehicle accidents: a systematic review, Transportation research part F: traffic psychology and behaviour 89 (2022) 249--264.

\bibitem{sharpe2013aesthetics}
S.~Sharpe, The aesthetics of urban movement: Habits, mobility, and resistance, Geographical Research 51~(2) (2013) 166--172.

\bibitem{lattarulo2019resistance}
P.~Lattarulo, V.~Masucci, M.~G. Pazienza, Resistance to change: Car use and routines, Transport Policy 74 (2019) 63--72.

\bibitem{douglas2024high}
K.~E. Douglas, A.~M. Fine, High speed, high risk: the rise of e-scooter injuries in adolescents, Pediatric Research (2024) 1--2.

\bibitem{guo2023psycho}
X.~Guo, A.~Tavakoli, A.~Angulo, E.~Robartes, T.~D. Chen, A.~Heydarian, Psycho-physiological measures on a bicycle simulator in immersive virtual environments: How protected/curbside bike lanes may improve perceived safety, Transportation research part F: traffic psychology and behaviour 92 (2023) 317--336.

\bibitem{MAD-IVE}
S.~Sabeti, A.~Tavakoli, A.~Heydarian, O.~Shoghli, Mad-ive: Multi-agent distributed immersive virtual environments for vulnerable road user research—potential, challenges, and requirements, in: Computing in Civil Engineering 2023, American Society of Civil Engineers (ASCE), 2024, pp. 1113--1120.

\bibitem{SABETI2024104867}
S.~Sabeti, F.~B. Ardecani, O.~Shoghli, Augmented reality safety warnings in roadway work zones: Evaluating the effect of modality on worker reaction times, Transportation Research Part C: Emerging Technologies 169 (2024) 104867.
\newblock \href {https://doi.org/https://doi.org/10.1016/j.trc.2024.104867} {\path{doi:https://doi.org/10.1016/j.trc.2024.104867}}.

\bibitem{Sabeti02012024}
N.~M. Sepehr~Sabeti, O.~Shoghli, Mixed-method usability investigation of arrows: augmented reality for roadway work zone safety, International Journal of Occupational Safety and Ergonomics 30~(1) (2024) 292--303.
\newblock \href {https://doi.org/10.1080/10803548.2023.2295660} {\path{doi:10.1080/10803548.2023.2295660}}.

\bibitem{BANANIARDECANI2025106802}
F.~{Banani Ardecani}, A.~Kumar, S.~Sabeti, O.~Shoghli, Neural correlates of augmented reality safety warnings: Eeg analysis of situational awareness and cognitive performance in roadway work zones, Safety Science 185 (2025) 106802.
\newblock \href {https://doi.org/https://doi.org/10.1016/j.ssci.2025.106802} {\path{doi:https://doi.org/10.1016/j.ssci.2025.106802}}.

\end{thebibliography}





\end{document}